# Active Heat Transfer Fluids (AHTF): Enhancement of Convective Heat Transfer by Bubble-Driven Self-Propelled Microparticles

Jacob Velazquez[1,2], Pawel Keblinski[3], Jeffrey L. Moran[1,4,*]

1 = Department of Mechanical Engineering, George Mason University, Manassas, VA

2 = Department of Physics, George Mason University, Manassas, VA

3 = Department of Materials Science and Engineering, Rensselaer Polytechnic Institute, Troy, NY

4 = Department of Bioengineering, George Mason University, Manassas, VA

*= Corresponding author. Email: jmoran23@gmu.edu

## Abstract

Liquid coolants containing conductive nanoparticles (nanofluids) have been widely studied over the past 30 years but have seen limited adoption in real-world cooling applications. The ability of passive nanoparticles to enhance heat transfer in liquids is fundamentally limited because the nanoparticles cannot move on their own relative to the bulk fluid, and thus generate negligible convective enhancements in heat transport. In this work, we present experimental evidence that micron-scale self-propelled particles, which convert chemical energy into autonomous motion, enhance convective heat transfer in liquids. We quantified this enhancement by measuring the convective heat transfer coefficient in a pool of suspension heated from below. The enhancements associated with self-propulsion are most pronounced at low heating powers (Rayleigh number of $\sim10^4$), in which case the heat transfer coefficient can be over 100% higher in the self-propelled case compared to the same particles without propulsion. This work provides a proof-of-concept demonstration that active particles can enhance heat transfer in liquids, motivating the development of "Active Heat Transfer Fluids" for various cooling applications.

## I. Introduction

Heat transfer fluids (HTF) are vital components of many of the engineering systems that make modern life possible. These include large data centers, graphical processing units, automobiles, chemical and paper processing, light-emitting diodes, and many others. Data centers alone account for between 1 and 1.5% of global energy consumption [1], and the carbon dioxide ($CO_2$) emissions associated with the information and communications industries are comparable to those of the airline industry [2]. This issue is exacerbated by the rapid growth of artificial intelligence (AI) and related technologies. For example, an average ChatGPT request requires about 10 times as much electricity to fulfill as an average Google search [3]. Enabling these technologies to advance while mitigating their negative environmental impacts thus depends on our ability to engineer more efficient HTF.

Liquid HTF make use of either single- or two-phase heat transfer. Two-phase heat transfer offers advantages such as exploiting liquids' high enthalpy of vaporization, but also presents unresolved safety and reliability challenges (e.g., the boiling crisis), as well as higher cost. In many applications, single-phase heat transfer remains the preferred option [4]. However, single-phase heat transfer also has limitations, including that most liquids that are safe to use as HTF, such as ethylene glycol or water, suffer from relatively low thermal conductivity, which limits their ability to dissipate heat.

One strategy to enhance HTF performance is to suspend thermally conductive nanoparticles in the coolant liquids, forming suspensions known as **nanofluids**. This approach was introduced in a seminal 1995 study [5] that triggered a flurry of research over the ensuing decades. Initially, several surprising results were reported, such as unexpectedly large nanoparticle-induced increases in thermal conductivity compared to the liquid alone [6,7], apparently defying effective-medium theories. Several hypotheses were proposed to explain these observations, such as that the Brownian motion of the nanoparticles agitates "micro-convection" that leads to the observed anomalous heat transfer enhancements [8]. However, this hypothesis is implausible considering that under typical conditions, the nanoparticles' diffusivity is orders of magnitude smaller than the thermal diffusivity of the liquid. This assertion is supported by molecular dynamics simulations [9,10] and carefully-executed experiments on nanofluids with well-dispersed nanoparticles [11,12]. In 2009, Buongiorno et al. published a benchmark study [13] on the thermal conductivity of nanofluids and found that Maxwell's effective-medium theory [14], generalized to account for interfacial resistance [15], accurately predicted the thermal conductivity of various nanofluids to within experimental uncertainty, suggesting that no anomalous heat transfer was present in these nanofluids [9,16]. This conclusion remains well-accepted today [17], and the early reports of anomalous thermal enhancements [6,7] are generally attributed to nanoparticle agglomerates through which heat preferentially flows, bypassing the liquid phase [18–23]. However, agglomeration-induced thermal conductivity enhancements have no practical value for HTF as they lead to increases in nanofluid viscosity [24,25]. The limitations on nanofluids can be traced, in part, to the fact that conventional nanoparticles cannot move autonomously and thus cause negligible convective enhancement [9,10,26].

Over the last twenty years, an intriguing development in nanotechnology has been the development of synthetic micro- and nanoparticles that propel themselves through liquids using energy they obtain from their environment. These **self-propelled particles (SPPs)** were first demonstrated in 2004 in the form of 2-µm-long, half-platinum half-gold (Pt/Au) nanorods that move with the Pt end forward in aqueous hydrogen peroxide ($H_2O_2$) solutions [27]. Since then, several SPP designs have been developed that use various fuel sources and move by different propulsion mechanisms [28]. SPPs may be considered the synthetic analogues of motile microorganisms and single cells, which play many important roles in biology [29].

As they move, both living and nonliving microswimmers generate disturbance flows that can enhance mixing in the surrounding fluid. In the early 2000s, the locomotion of **Escherichia coli** bacteria was reported to lead to enhanced mixing of passive tracers [30,31]. It has since been established both theoretically [32,33] and experimentally [34] that the extent of SPP-induced mixing depends on the details of the locomotion mechanism. In particular, microswimmers that are propelled from behind, often called "pushers" (such as *E. coli*) cause strong mixing and active turbulence [32,34], while those propelled from the front, known as "pullers" (such as the marine alga *Chlamydomonas reinhardtii*), do not induce significant mixing, akin to passive particles [32]. Intriguingly, suspensions of pushers can also exhibit lower viscosity than the fluid alone [35,36], owing to the active stresses exerted on the fluid by the swimmers and the hydrodynamic interactions among them, in some cases leading to rheological behavior resembling superfluids [36]. In contrast, pullers lead to an increase in the overall viscosity, in the same manner as (but to a greater extent than) passive particles [37]. The rheological properties of active suspensions were reviewed by Saintillan [38].

In addition to mass and momentum, the convective disturbances caused by microscale motile cells or particles can also lead to enhanced mixing of thermal energy. In 2013, Solis and Martin demonstrated that the heat transfer rate in a suspension of externally-driven magnetic platelets (~20 μm wide and ~1 μm thick) can be controllably varied by driving them in different trajectories [39]. By applying different direct or alternating current (DC or AC) magnetic field waveforms, they found that the effective thermal conductivity of the suspension could be varied over several orders of magnitude, from nearly zero to over 18 W m$^{-1}$ K$^{-1}$ (comparable to that of stainless steel). Similar findings have been reported in natural systems: in a meromictic lake (i.e., one whose layers do not naturally intermix), the motions of *Chromatium okenii* bacteria generated thermal mixing that was significant enough to homogenize the fluid temperature across a depth of roughly 1 m, which is approximately 5 orders of magnitude larger than an individual bacterium [40]. In a separate study involving the engineered fluid HFE-7000 in water, the gravity-driven rising and falling motions of microscale HFE-7000 bubbles and droplets, respectively, were shown to generate turbulent mixing and produce a fivefold increase in the convective heat transfer rate compared to the case of classical thermal turbulence [41,42]. These previous studies demonstrate the potential of microscale motion (driven by magnetic fields, biological motility, or buoyancy forces) to drive convective thermal mixing in liquids. However, this principle has yet to be demonstrated experimentally with autonomously motile SPPs. SPPs have the advantage that they do not require any external driving, instead generating their own motion through various mechanisms (e.g., chemical reactions on the particle's surface that lead to propulsive forces).

Only a handful of studies have analyzed the potential cooling applications of SPPs. In 2017, El Hasadi et al. proposed a "self-propelled nanofluid" consisting of carbon-nanotube-based SPPs [43], predicting that self-propulsion could more than triple the heat transfer coefficient compared to the same suspension without self-propulsion. In addition to heat transfer enhancement, if the SPPs move as "pushers" (similar to *E. coli*), El Hasadi et al. predicted that the SPPs' motion could also reduce the suspension's viscosity by a factor of more than 25 compared to the liquid alone, following the aforementioned reports of viscosity reduction by pullers [35,36]. In a follow-up study [44], El Hasadi et al. theoretically analyzed the dependence of thermal mixing enhancement on the SPPs' speeds and aspect ratios, finding that larger SPPs, faster speeds, and larger aspect ratios yield the most enhancement. Finally, in our previous work, Peng et al. modeled the effects of self-propelled nanoparticles on thermal transport through a model liquid using molecular dynamics simulations [45]. For a constant heat flux through the fluid, we found that the steady-state temperature gradient is shallower in the presence of self-propulsion compared to the case of pure Brownian motion, implying that self-propulsion increases the suspension's "effective thermal conductivity" by roughly 10% compared to the same nanofluid without propulsion. This increase is significant considering the small size of the nanoparticles: indeed, the equivalent physical size in this case is on the order of only 5 nm. Most SPPs are significantly larger, ranging in size from 0.1 to 100 µm, so based on the theoretical modeling of El Hasadi et al. [44], it is expected that the increases could be significantly greater for larger particles, faster speeds, and higher volume fractions.

In summary, there is substantial theoretical and experimental evidence that SPP suspensions could enable transformative improvements in liquid coolant technologies. However, there remains a lack of experimental demonstration of SPP-induced heat transfer enhancement using self-propelled colloids that do not require external forcing. In addition, experimental data quantifying the extent to which SPPs enhance the convective heat transfer coefficient (HTC), a widely accepted metric of heat transfer enhancement, has never been reported. As a result, there is a knowledge gap related to the lack of experimental data confirming the ability of SPPs to enhance the mixing of heat controllably in real fluids.

In this work, we fill this knowledge gap by providing experimental data showing enhancement of convective heat transfer by artificial SPPs, specifically bubble-propelled manganese dioxide ($MnO_2$) microparticles. We use two metrics to quantify the impact of particle self-propulsion on heat transfer. First, we visualize SPP-induced disturbances in the fluid's temperature distribution using infrared thermography, demonstrating qualitatively that they outpace thermal diffusion (in contrast to the purely Brownian motion of the particles in conventional nanofluids [9]). Second, we quantify HTC enhancement in an SPP suspension heated from below, finding increases of up to 100% associated with self-propulsion. These increases are most significant at the lowest Rayleigh numbers (corresponding to minimal heat input) and decrease as the heating power (and thus Rayleigh number) increases. This work demonstrates a proof of principle that SPPs can exert a significant influence on heat transfer in liquids It also opens the door for numerous follow-up studies of this phenomenon, such as its dependence on SPP size, shape, propulsion mechanism, and volume fraction, toward further engineering of these "active heat transfer fluids" (AHTF) for a variety of high-performance cooling applications.

The article is organized as follows. Section II describes our methodology for the two sets of experiments performed. Section III shows results from the first set of experiments, in the form of IR images that visualize SPP-induced disturbances in the temperature distribution, motivating the idea that they can enhance convective heat transfer. Section IV shows our results for the heat transfer coefficient in SPP suspensions. In Section V, we provide a discussion of the results. Finally, in section VI, we offer concluding thoughts and future plans.

## *II.  Methods*

The SPPs used herein were manganese dioxide ($MnO_2$) microparticles, which propel themselves in aqueous solutions of hydrogen peroxide ($H_2O_2$) by catalytically decomposing $H_2O_2$ into water and oxygen ($O_2$) bubbles, as was first demonstrated in ref. [46]. This design was selected for two reasons. First, bubble-propelled SPPs are relatively efficient at agitating convective mixing, which has been demonstrated previously [47], while other mechanisms, such as autophoresis [48], have relatively low efficiency because of their slower propulsive speeds [49]. (The $MnO_2$ SPPs were previously shown to achieve speeds faster than 100 µm s$^{-1}$ at fuel concentrations below 1 wt.% [46].) Second, $MnO_2$ microparticles are convenient since they can be purchased commercially and used with minimal preparation. Owing to the irregular shapes and surface texture of the $MnO_2$ particles (Figure 1A), $O_2$ bubbles nucleate and detach at random locations on the surface of each particle, and this symmetry breaking leads to propulsion in randomly-varying directions. Upon detachment, the $O_2$ bubbles rise buoyantly, and this rising motion induces additional mixing of momentum and thermal energy.

Figure 1A shows a scanning electron micrograph of the MnO$_2$ microparticles that were used in this study, with an inset showing a zoomed-in view of a single particle. Figure 1B shows a screen capture of a microscope video (provided in the Supplemental Information) showing the self-propelled motion of the MnO$_2$ particles. The "trains" of bubbles behind individual MnO$_2$ SPPs are visible.

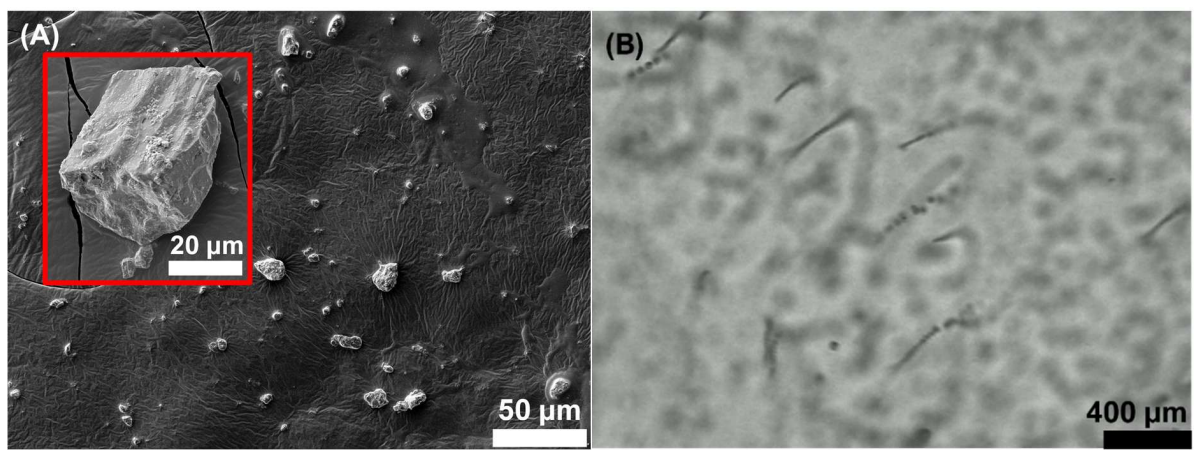

**Figure 1: (A) Scanning electron micrograph of manganese dioxide (MnO$_2$) microparticles; scale bar = 50 μm. Inset shows an individual particle (scale bar in inset = 20 μm). (B) Screen capture from microscope video showing several MnO$_2$ SPPs moving in an H$_2$O$_2$ solution, as well as several out-of-focus oxygen bubbles generated by the particles, which rise buoyantly upon detachment. Scale bar = 400 μm.**

The MnO$_2$ particles were purchased commercially (Loudwolf, product LW-MNO2-16/1, Dublin, CA, USA) and used as received. They show polydispersity in their size and morphology. Figure 2 shows the distribution of the maximum Feret diameters of 1,560 particles, as determined through analysis of optical micrographs (included in the supplementary information). Figure 3 shows the distribution of aspect ratios of the particles, defined as their maximum Feret diameter divided by minimum Feret diameter. As this figure shows, although a few large particles are depicted in Figure 1, most particles are smaller, on the order of 2 to 5 μm. Additionally, Figure 3 demonstrates that the vast majority of the particles characterized have an aspect ratio between 1 and 2. In this study, for simplicity, we used the particles as received with no filtration steps, and we only used one volume fraction and one fuel concentration. A detailed assessment of the effect of particle size, shape, and volume fraction on heat transfer enhancement is beyond the scope of the present work.

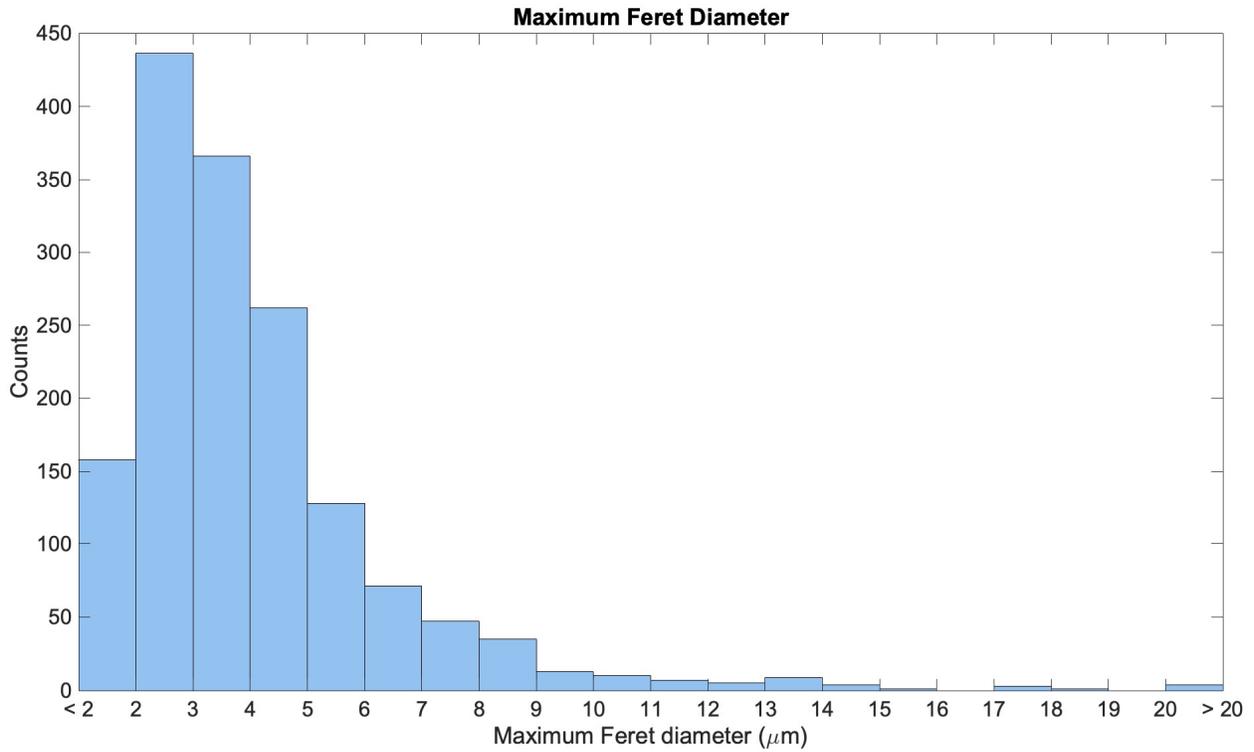

**Figure 2: Distribution of the maximum Feret diameters of a sample of 1,560 manganese dioxide (MnO$_2$) particles. The data was obtained from brightfield microscope images, included in the SI, and processed using ImageJ v2.16 software.**

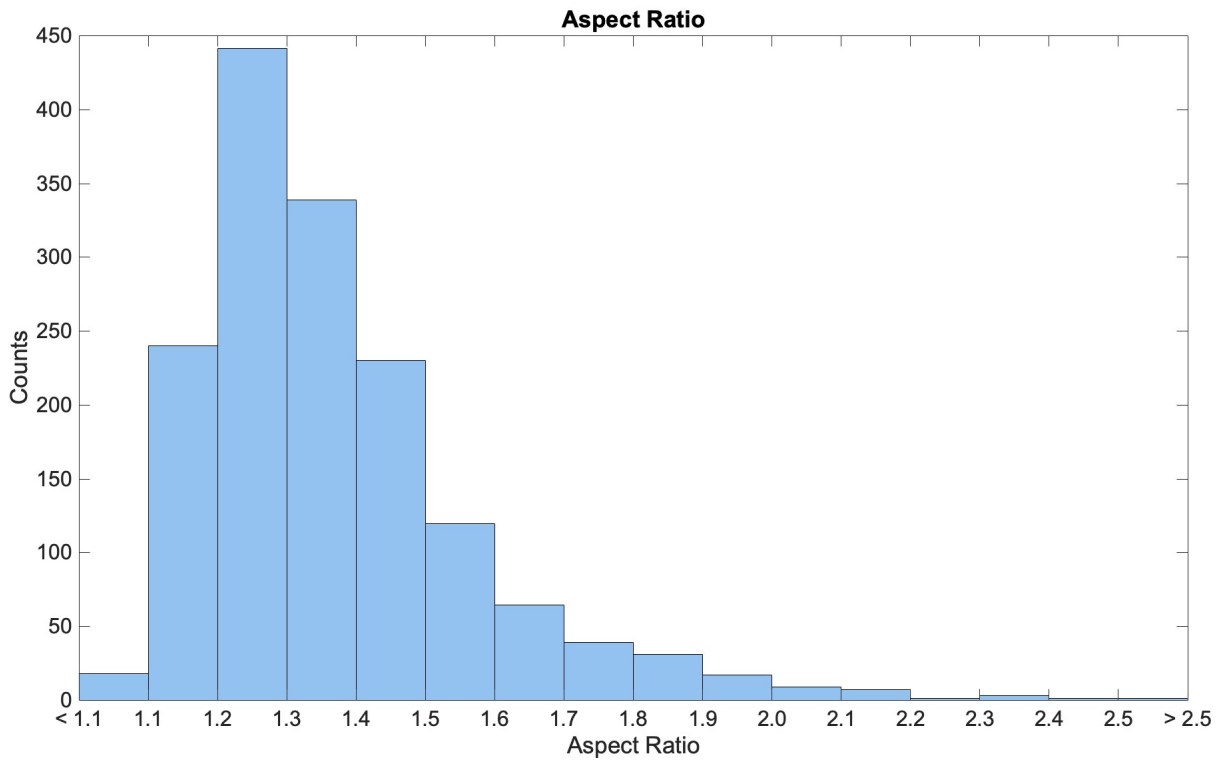

**Figure 3: Distribution of the aspect ratios of a sample of 1,560 manganese dioxide (MnO$_2$) particles. The data was obtained from brightfield microscope images, included in the SI, and processed using ImageJ v2.16 software.**

As with most chemically-fueled SPPs, the speed of MnO$_2$ SPPs increases as the H$_2$O$_2$ fuel concentration is increased [46]. In this work, for simplicity an H$_2$O$_2$ concentration of 0.3 wt. % was used for all experiments. To facilitate detachment of O$_2$ bubbles from the particles' surfaces and to maintain stability of the colloidal suspension, a surfactant was added to the solution. The important role of surfactants in facilitating bubble-propelled SPPs' motion was explored by Wang et al. [50]. Throughout this work, we used Triton X-100 at a constant concentration of 0.5 wt. %. An important future study would seek to quantify the effect of fuel concentration (and thus of self-propelled speed) on the heat transfer enhancement. However, such studies would be complicated by the inevitable variations in speed from one particle to another, even at a constant fuel concentration. Since in this work our primary objective is to demonstrate the concept of SPP-induced heat transfer enhancement, herein we keep $H_2O_2$ concentration fixed at a relatively modest 0.3 wt. %, and compare the self-propulsion case (in which $H_2O_2$ is present and propulsion occurs) to the control case of identical particles and surfactant, but no $H_2O_2$ fuel, in which case the motion of the $MnO_2$ microparticles is governed by Brownian motion and gravitational settling.

We ran two sets of experiments, each of which quantifies thermal mixing by different metrics. The first set of experiments visualized the SPP-induced perturbations to the temperature distribution in the liquid, which was done using the apparatus shown in Figure 4. A Peltier heater is positioned vertically in a glass Petri dish that contains 8 mL of liquid, corresponding to a relatively thin liquid film (~4 mm deep). A constant current and constant voltage are supplied to the heater, resulting in an approximately constant heat transfer rate to the fluid. The liquid film is kept thin for two reasons: (1) to minimize temperature variations in the vertical direction, allowing us to approximate the temperature distribution as two-dimensional (2D); and (2) to keep the Rayleigh number (using the film depth as a length scale) below the approximate threshold of ~$10^4$ above which natural convection flows become significant in this geometry [51]. This setup allows visualization of the effects of SPP-induced convection without the confounding influence of natural convection.

The temperature distribution in the liquid film was visualized using a Teledyne FLIR (Wilsonville, OR) E8-XT infrared camera. Images were acquired manually at 30-s intervals. For the temperature range considered (20 to 40°C), we assumed that the emissivity $\varepsilon$ of the free surface of the liquid was equal to that of liquid water, assumed to be 0.97 in all cases [52]. Within the spectral range of the camera (between 7.5 and 13 µm), the penetration depth of water is on the order of 20 µm [53]. Thus, the temperature distributions in Figure 6 may be assumed to represent an average temperature over the fluid within approximately twice the penetration depth (i.e., about 40 μm) of the free surface. Although the temperature distribution in the fluid is not perfectly 2D, this methodology allows us to fulfill our objective of qualitatively visualizing SPP-induced disturbances to the temperature distribution in the fluid.

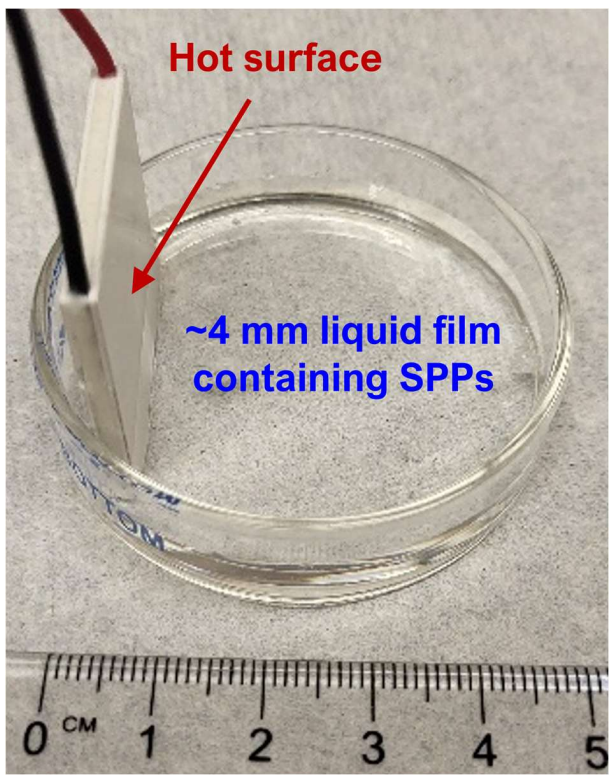

**Figure 4: Experimental setup for IR imaging experiments. A Peltier heater is positioned vertically in a Petri dish containing the SPP suspension. The liquid film is approximately 4 mm thick, which ensures that the temperature distribution is approximately two-dimensional (2D) and that natural convection is minimized. As the Peltier heater supplies heat to the fluid at a constant rate, the temperature distribution is recorded from above using an infrared camera (not shown).**

The purpose of the second set of experiments was to quantitatively assess the impact of self-propulsion on heat transfer in the suspension, specifically by determining the heat transfer coefficient (HTC). The setup for these experiments is shown in Figure 5 and consists of a cubical fluid container (~16 cm on one side; see item 1 in Figure 5) into which approximately 3 L of SPP suspension is poured. A resistive heat source (item 2 in Figure 5) at the bottom injects heat into the fluid, and the heat flux $q''$ and temperature $T_s$ at the heated surface are recorded by a thermopile and T-type thermocouple, respectively. The thermopile and thermocouple are packaged together in a Hukseflux (Delft, Netherlands) FHF05-15X30 sensor glued to the heater surface (see item 3a in Figure 5). The thermopile sensors have a nominal sensitivity of $3 \times 10^{-6}$ V (W m$^{-2}$)$^{-1}$ and a measurement uncertainty of ±5% ($k = 2$). The factory calibration method follows the recommended practice of ASTM C1130-21.

At the free surface of the SPP suspension, a second Hukseflux FHF05-15X30 sensor records the temperature, which we assume to be the "background" fluid temperature, $T_\infty$. This location is far enough away from the heater that it is nearly unaffected by the heat flux into the fluid, so it can be used to compute the heat transfer coefficient by rearranging Newton's Law of Cooling:

$$h = \frac{q''}{T_s - T_\infty} \qquad (1)$$

Here, $q''$ and $T_s$ are recorded by the thermopile and thermocouple within the sensor mounted on the heat source (item 3a in Figure 5) and $T_\infty$ is recorded by the second thermocouple (item 3b in Figure 5). Measurements of $q''$, $T_s$, and $T_\infty$ are automatically gathered at 30-s intervals over the course of the experiment. By inserting these values into the right-hand side of equation (1), we can solve for $h$ at the same times.

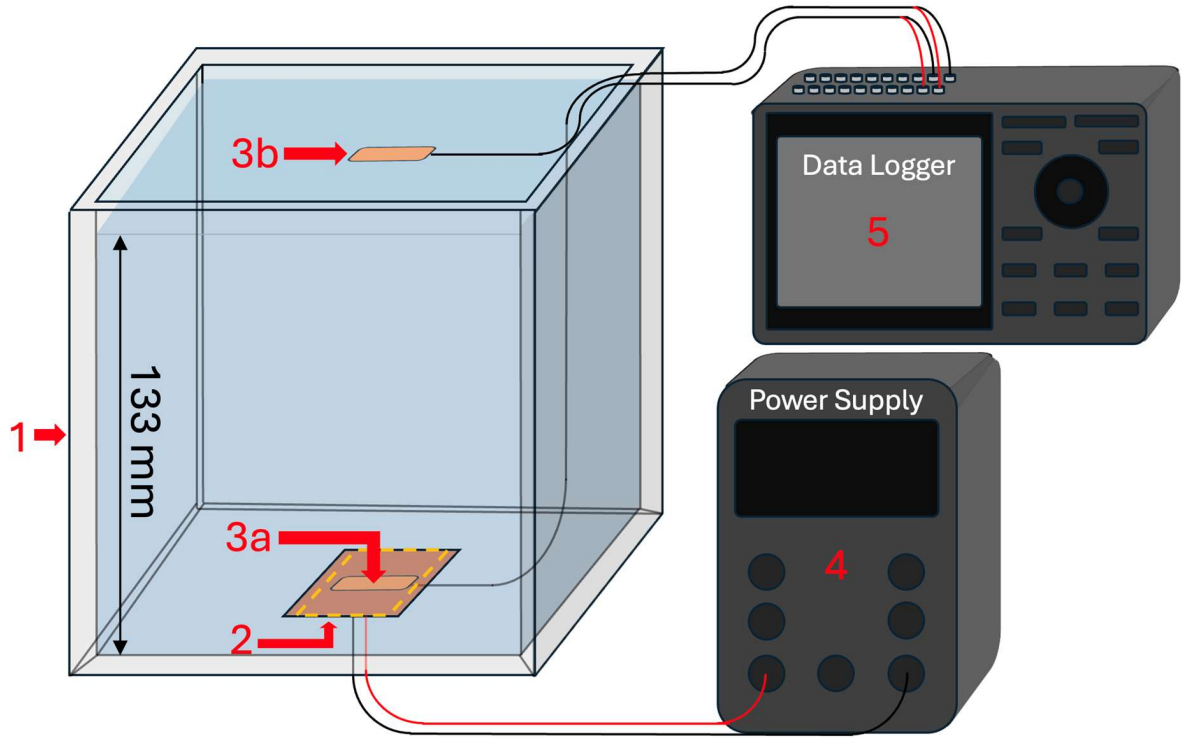

**Figure 5: Experimental setup to quantify the convective heat transfer coefficient (HTC) for suspensions of self-propelled particles (SPPs). (1) A cubical "fish tank" (160 x 160 x 160 mm³) contains 3 L of the SPP suspension, resulting in a liquid depth of 133 mm. (2) A copper heat spreader with a square shape (4 cm on each side) is glued to the bottom inside the fish tank and a resistive heat source (3 x 4 cm²) is glued to the underside of the fish tank to supply heat to the SPP suspension; its position is indicated by the gold dashed box. (3) Two Hukseflux FHF05-15X30 sensors can measure temperature and heat flux simultaneously. Sensor (3a) records the temperature $T_s$ and heat flux $q''$ at the heated surface; sensor (3b) sits near the free surface of the fluid and reports the background temperature, $T_\infty$. (4) NANKADF power supply, which supplies a constant voltage and current to the resistive heater. (5) Graphtec data logger to record raw temperature and heat flux data.**

Prior to each experiment, 3 L of deionized water and 15 mL (0.5 vol. %) of Triton X-100 were added to our 1-gallon cubical "fish tank" container. Then the $MnO_2$ particles were added to the fluid and dispersed using the micro-tip of a Bransonic probe sonicator (model SFX150; Brookfield, CT, USA) for 10 minutes at 70% amplitude. After that, each experiment consisted of two datasets gathered in immediate succession. First, we ran a control experiment in which the aqueous suspension of $MnO_2$ microparticles and Triton X-100 without $H_2O_2$ fuel was subjected to a constant heat input, and the heat fluxes and temperatures were recorded for 30 minutes. Next, the suspension was once again sonicated (to disperse the particles, some of which sediment to the bottom during the control experiment), $H_2O_2$ fuel was mixed into to the solution, and the measurements were repeated, again for 30 minutes, now in the presence of self-propulsion. This approach – running the self-propelled case immediately after the control case – was chosen for three reasons. First, it helped conserve resources, since each experiment uses a relatively large quantity of particles and chemicals. Second, it enables a meaningful comparison between the control case and the self-propelled case, since we are comparing the HTC for the exact same particle suspension before and after the addition of $H_2O_2$ fuel. Third, it avoids complications due to trial-by-trial variation in the ambient experimental conditions (such as the background air temperature or the fluid temperature), isolating only the effects of $H_2O_2$ addition and the resultant self-propulsion on the HTC.

Despite these precautions, there remained minor variations in the initial conditions of each trial. For example, the initial velocity distribution of the microparticles varied from one experimental trial to the next due to the mixing of the hydrogen peroxide into the control suspension. Further, while the precise temperature distribution within the fluid was similar at the beginning of each trial of each heating condition, it varied by the heating condition, since a longer cooling time between the control and the self-propelled cases would have allowed re-aggregation and sedimentation of the particles. For these reasons, we chose to omit the first ten minutes of data acquired in each experiment to allow the system to reach a "quasi-steady state" (acknowledging that since this system is inherently out of thermodynamic equilibrium, a true steady state will never be reached).

To understand the role of conduction heat transfer through the suspensions, we also quantified the thermal conductivity of both aqueous $H_2O_2$ solutions alone and the $MnO_2$ microparticle suspensions without $H_2O_2$. The thermal conductivity was measured using a TCi thermal conductivity analyzer (C-Therm, Fredericton, NB, Canada), using the "Liquids and Powders default" measurement setting.

For each power setting, a characteristic Rayleigh number was calculated based on the average temperature differential of each of the control experiments (i.e., no $H_2O_2$, so that natural convection was the only source of fluid motion) at that power. The Rayleigh number $Ra_{L_c}$ and associated length scale $L_c$ are defined, respectively, by

$$Ra_{L_c} = \frac{g\beta(T_s - T_\infty)L_c^3}{\nu\alpha}, \tag{2}$$

$$L_c = \frac{A_s}{P}, \tag{3}$$

where $T_s$ is the temperature of the heated surface, $T_\infty$ is the background fluid temperature (measured by sensors 3a and 3b, respectively, in Figure 5), the gravitational acceleration $g = 9.81$ m/s², and the thermal expansion coefficient of water is $\beta = 2.07 \times 10^{-4}$ K⁻¹ at 20°C and $3.03 \times 10^{-4}$ K⁻¹ at 30°C. For our calculations, the precise value of $\beta$ was evaluated at the film temperature (arithmetic mean of $T_s$ and $T_\infty$). The kinematic viscosity and thermal diffusivity of water were both evaluated at 25°C to be $\nu \approx 8.93 \times 10^{-7}$ m² s⁻¹ and $\alpha \approx 1.46 \times 10^{-7}$ m² s⁻¹, respectively [54,55].

We note that $L_c$ is an imperfect estimate of the characteristic length for Rayleigh number since both $A_s$ and $P$ grow over time as heat diffuses through the glass underside of the fish tank. For simplicity and to facilitate comparison among cases, we assume $L_c$ is fixed. This is a reasonable assumption since heat spreads through glass in roughly the same fashion for all trials. The choice of a constant $L_c$ ensures that facile comparisons can be made between different cases.

### III.     Results of Infrared Imaging Experiments

In the first set of experiments, we sought to quantify the temperature distribution in a fluid containing SPPs and compare said distribution with the control case of water only. The apparatus for these experiments is pictured in Figure 4. As mentioned above, prior literature establishes that Brownian motion of passive nanoparticles does not significantly affect heat transfer in the surrounding liquid [9–12], primarily because under typical conditions, the thermal diffusivity of the liquid is much larger than the Brownian diffusivities of the nanoparticles, and so thermal diffusion dominates over Brownian-motion-induced microconvection. (For instance, the thermal diffusivity of water at room temperature is about 3 orders of magnitude larger than that of 10-nm nanospheres.) Considering these findings, it was important to establish that SPPs can generate disturbances to the temperature distribution that can outpace thermal diffusion.

Since the propulsion direction of an unguided SPP typically varies in time, the motion of SPPs is often referred to as an "enhanced diffusion" process, often modeled as an "active Brownian particle" (ABP). The overall result of this is that the behavior macroscopically resembles diffusion, but with an enhanced diffusivity [56]. For an SPP that translates with a constant speed $U$, the swim diffusivity is quadratic in $U$ [38,56]. El Hasadi et al. argued that for SPPs to enhance heat transfer in liquids, their swim diffusivity should be comparable to or exceed the thermal diffusivity of the fluid [44]. As a result of these enhancements, the swim diffusivity can exceed the Brownian diffusivity by more than three orders of magnitude [56], and thus is capable in principle of exceeding the thermal diffusivity of the liquid. The first set of experiments sought to verify this prediction experimentally.

Figure 6 shows infrared (IR) images of the temperature distribution in the fluid in the cases of pure water and MnO$_2$ SPPs, respectively. Considering that the voltage (2.4 V) and current (0.5 A) supplied to the Peltier heater are the same in both cases, the total power supplied to the heater is the same in both cases. In each image, the temperature in crosshairs is indicated in the upper-left corner of the image. In the MnO$_2$ case, the starting temperature is warmer (24.3°C) at the initial time than the control (22.7°C). This is because of the ultrasonication process that was performed on the MnO$_2$ suspension, whose purpose is to disperse the microparticles, generates waste heat that raises the temperature of the suspension. This elevated initial temperature should be borne in mind when interpreting the temperature data. Nevertheless, the temperature difference between the center of the dish and the heater surface is lower in the presence of MnO$_2$ SPPs after 1, 5, or 10 minutes than for water alone. Over the course of 10 minutes, the fluid in the center of the dish experiences a larger temperature rise in the MnO$_2$ SPP case (increasing by 6.7°C, from 24.3 to 31.0°C) compared to the control case (increasing by 4.1°C, from 22.7 to 26.8°C). These observations suggest, but do not demonstrate definitively, that in the presence of SPPs, thermal energy is extracted from the heater at a faster rate compared to the case of water alone, which would manifest as an increase in the overall *average* temperature of the fluid. These observations led us to hypothesize that the convective heat transfer coefficient (HTC) is larger in the presence of SPPs than in their absence. To test this hypothesis quantitatively, we developed a separate experimental setup to measure the convective heat transfer coefficient in the presence of self-propelled particles.

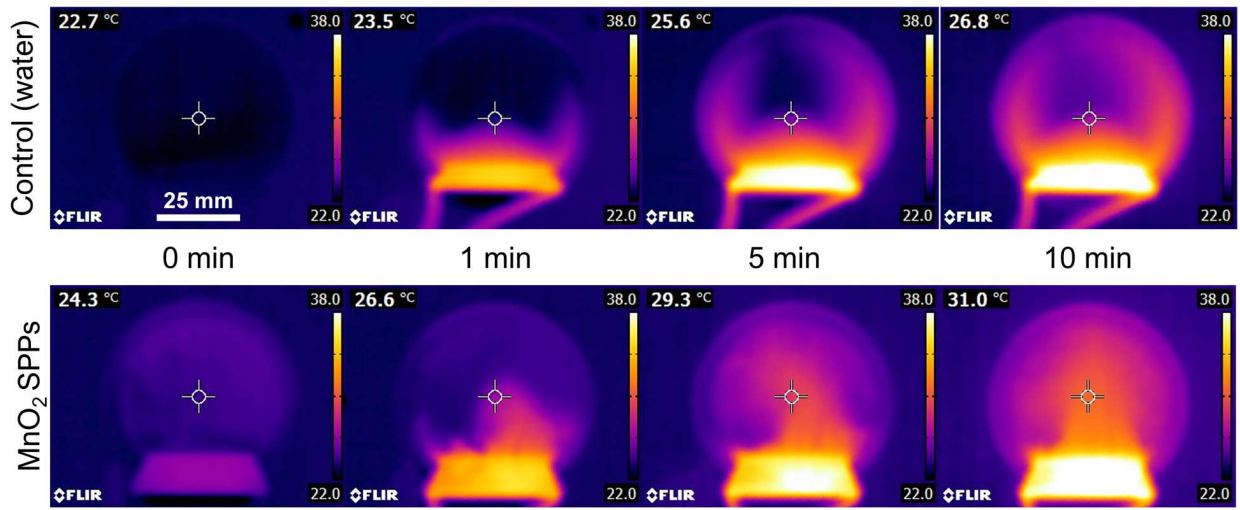

**Figure 6: Temperature distribution (visualized by IR thermography) in water (top row) and an MnO₂ SPP suspension (bottom row), using the apparatus pictured in Figure 4. In the control case, the fluid is water and the temperature distribution evolves smoothly in space and time, indicating that conduction is the primary mode of heat transfer. In the self-propelled case (bottom row), the fluid is an aqueous suspension of MnO₂ SPPs in 0.3 wt. % H₂O₂ with 0.5 wt. % Triton X-100, and the temperature profile is more irregular. In both conditions, the power supplied to the heater is approximately 1 W. The color scale is the same for all figures. At each time point, the temperature in the center of the image (in crosshairs) is indicated in the top-left of the respective panel. Comparison of the two cases at the same elapsed time qualitatively illustrates the effect of SPPs on thermal mixing. Moreover, it demonstrates that the disturbance flows generated by the MnO₂ SPPs, as well as the O₂ bubbles they generate, outpace thermal diffusion in the fluid. The scale bar (bottom of top-left image) represents 25 mm.**

## IV.   Heat Transfer Coefficient Measurements

To quantify the enhancements to heat transfer resulting specifically from the motion of the MnO₂ SPPs, as well as the bubbles they generate, it is important to rule out increases from other sources, such as an increase in the thermal conductivity of the solution resulting from the addition of MnO₂ microparticles, Triton X-100 surfactant, or H₂O₂ fuel. Thus, before presenting the HTC enhancement results, it is important to determine the effects of the addition of $MnO_2$ microparticles, $H_2O_2$, and Triton X-100 on the thermal conductivity of the solution.

**Table 1: Thermal conductivities (W m$^{-1}$ K$^{-1}$) measured with the modified transient plane source method (MTPS). Each entry represents the average of ten (10) measurements $\pm$ one standard deviation.**

|  | Pure Water | Suspension of MnO2 (1 mg/mL) and Triton X-100 (0.5%) in water | Hydrogen Peroxide (0.3%) in water | Hydrogen Peroxide (30%) in water |
|---|---|---|---|---|
| k (W m$^{-1}$ K$^{-1}$) | 0.6064 ± 0.0038 | 0.6001 ± 0.0018 | 0.6018 ± 0.0022 | 0.5848±0033 |

Although it would be ideal to quantify the thermal conductivity of a suspension of $MnO_2$ particles in an aqueous solution of surfactant and $H_2O_2$, such measurements would be practically impossible since the $MnO_2$ particles propel themselves through the liquid, causing convective disturbances (which we seek to quantify in this section), obscuring the contributions from conduction alone. Recognizing this, we sought to quantify the contributions of $H_2O_2$ and surfactant separately from those of suspensions of the $MnO_2$ particles. Accordingly, **Table 1** shows the thermal conductivities of pure deionized water, an $MnO_2$ suspension in water and surfactant, and aqueous solutions of hydrogen peroxide at 0.3% and 30% concentrations. **Table 1** indicates that the addition of $H_2O_2$ to water lowers its thermal conductivity and that the addition of $MnO_2$ particles and surfactant also produces a slight decrease in thermal conductivity compared to water alone. Taken together, the data in **Table 1** indicate that any enhancement in heat transfer observed in the experiments cannot result from an increase in the suspension thermal conductivity.

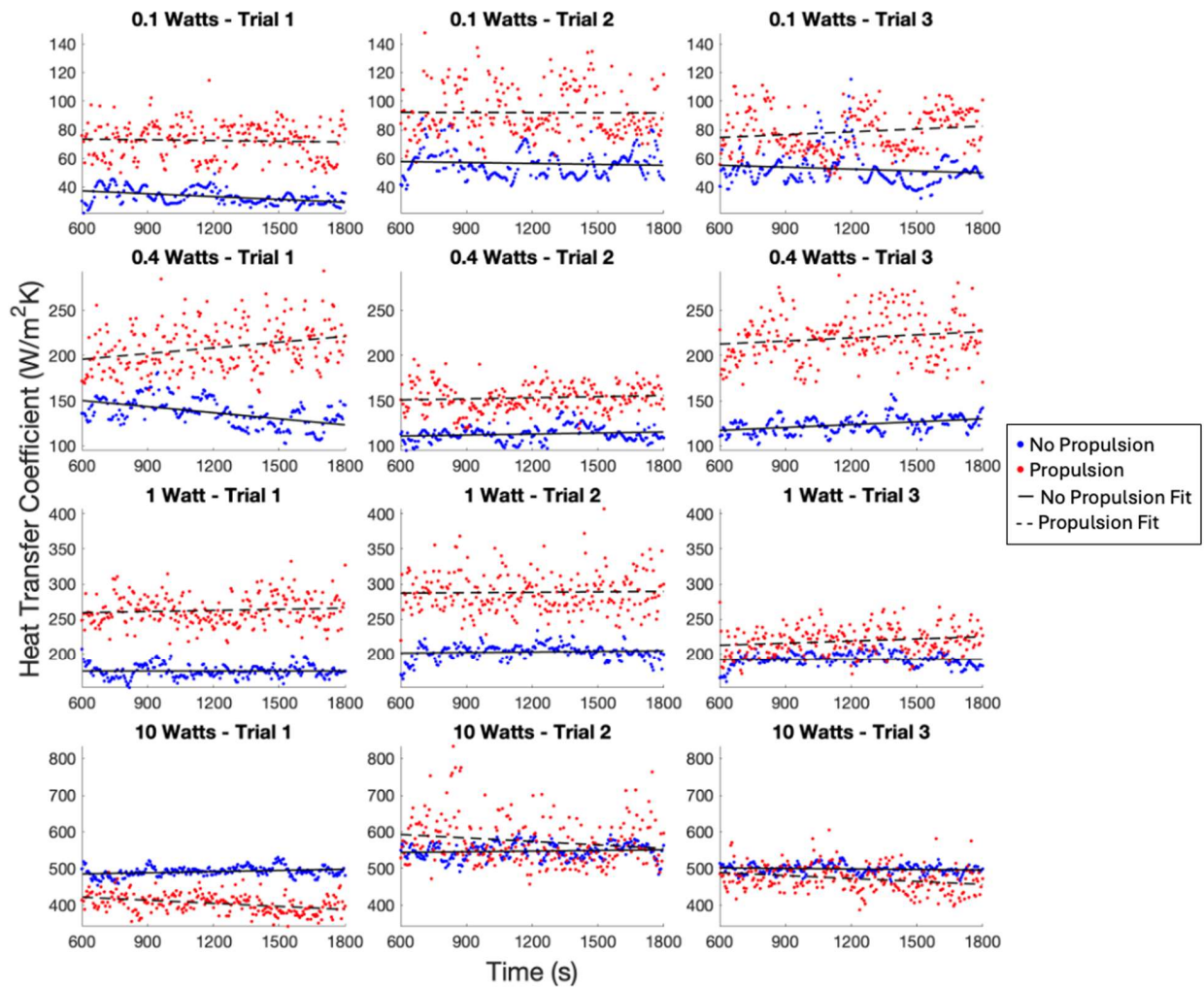

**Figure 7:** Heat transfer coefficient (HTC) as a function of time (from 10 to 30 minutes into the experiment) for four different rates at which power is supplied to the heater (0.1, 0.4, 1, or 10 W). Columns correspond to individual experiments at each heating power. Transient behavior from $0 < t < 600$ s varies from trial to trial and is omitted for clarity. In each panel, the blue data ("No Propulsion") indicates the case of an aqueous suspension of manganese dioxide ($MnO_2$) microparticles and Triton X-100 with no $H_2O_2$ fuel. The red data ("Propulsion") indicates the case in which hydrogen peroxide fuel (0.3 wt.%) has been added to the suspension, causing the SPPs to propel themselves through the fluid. Solid and dashed lines show linear least-squares fits to the control and self-propulsion data, respectively.

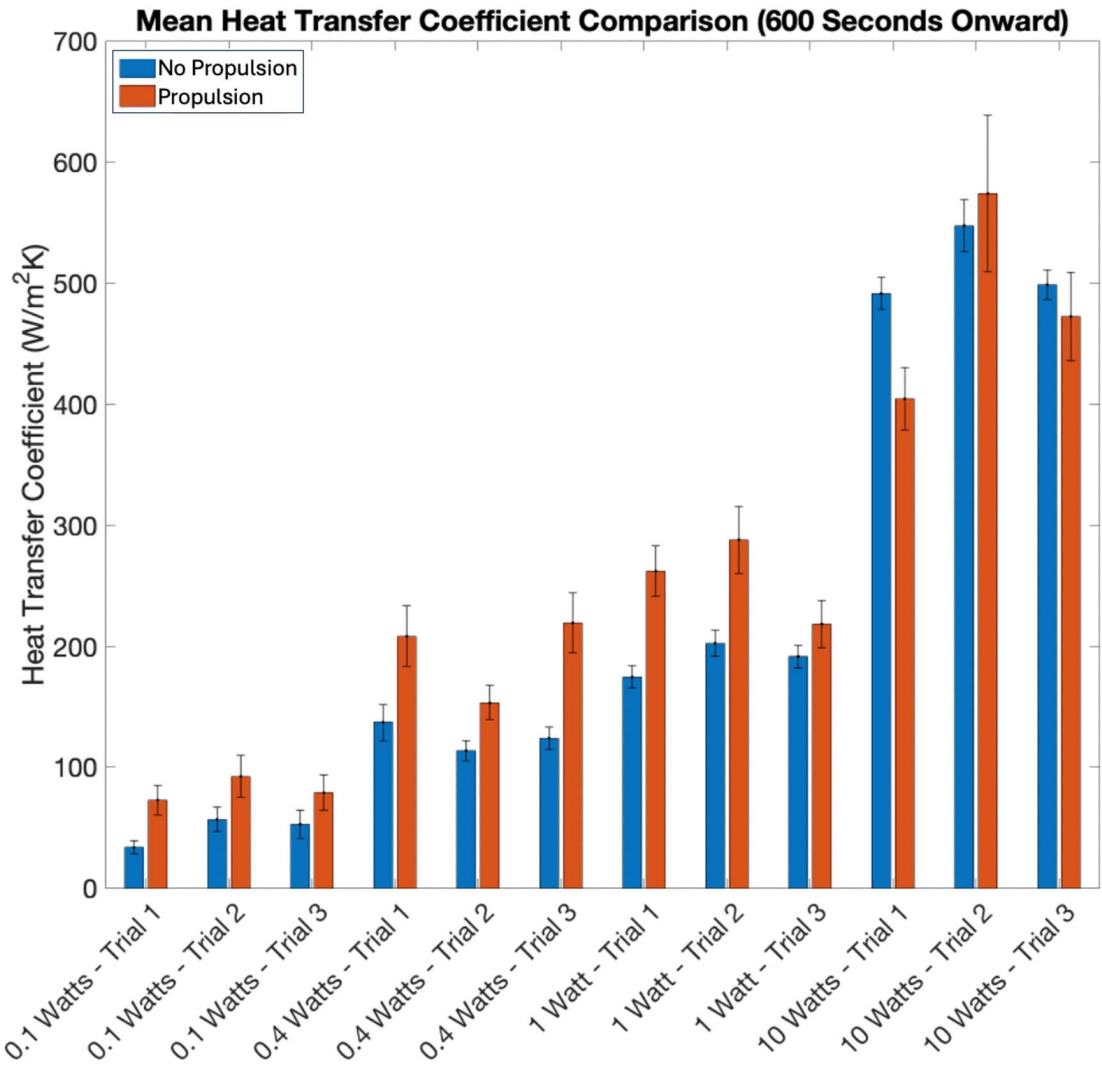

**Figure 8: Average heat transfer coefficient (HTC) values for all experimental trials. Data are averages computed from the data shown in from Figure 7; transient effects from $0 < t < 600$ s were not considered. Error bars show one standard deviation in each direction. To capture a "quasi-steady state" in which the HTC fluctuates about an approximately steady mean value, the data shown are averages over the time period from $t = 600$ s to $t = 1800$ s.**

**Figure 7** shows experimental data for the convective heat transfer coefficient (HTC) (computed from the heat flux and temperature data via Newton's Law of Cooling, equation (1)) in an SPP suspension in the presence and absence of particle self-propulsion. There is significant enhancement of the HTC in the presence of self-propulsion ("Propulsion") compared to the control case ("No Propulsion"), especially at low heating powers. In the control case, we include the very same MnO$_2$ particles and surfactant, but no H$_2$O$_2$.

For each experimental trial, we show the results from $t = 600$ s to $t = 1800$ s. The transient results ($t = 0$ to $t = 600$ s) are omitted. This transient phase is ignored because, as mentioned above, the active matter system considered here is inherently out of thermodynamic equilibrium owing to the continuous dissipation of energy by the SPPs, and so a true steady state cannot be reached while self-propelled motion is occurring. To estimate the HTC enhancement associated with SPPs, we define a "quasi-steady state" (assumed to begin at $t = 600$ s) in which the values of the HTC are assumed to fluctuate about a mean value. As demonstrated by the near-horizontal least-squares fit lines in **Figure 7**, the system attains a quasi-steady state during this period in the sense that the HTC is approximately constant in time. **Figure 8** and **Table 2** depict the numerical values of the HTC for each of the trials depicted graphically in **Figure 7** (shown as arithmetic means of the HTC measurements over the time interval $600\ s < t < 1800\ s$ ± one standard deviation).

Since the MnO$_2$ microparticles are propelled by catalytically-generated O$_2$ bubbles, the enhancements in heat transfer coefficient are attributed both to the self-propelled motion of the particles and to the rising motions of the bubbles themselves, which rise from buoyancy upon detachment from the particles. Understanding the relative impact of the bubbles vs. the particles is difficult, and we do not attempt to do so here. We provide further commentary on this point in §V, along with recommendations to disentangle these contributions in the future.

**Table 2: Heat transfer coefficient (W m⁻² K⁻¹) for the data depicted in Figure 7. Each entry represents the steady-state time-averaged value for each trial ± one standard deviation.**

| Power Input (W) | Trial 1 | | Trial 2 | | Trial 3 | |
|---|---|---|---|---|---|---|
| | No Propulsion | Propulsion | No Propulsion | Propulsion | No Propulsion | Propulsion |
| 0.1 | 33.56 ± 5.38 | 72.71 ± 12.12 | 56.69 ± 9.98 | 92.17 ± 17.1 | 52.76 ± 11.61 | 78.7 ± 14.6 |
| 0.4 | 136.92 ± 15.07 | 208.63 ± 25.11 | 113.27 ± 8.14 | 153.5 ± 14.62 | 123.75 ± 9.25 | 219.65 ± 24.9 |
| 1 | 174.92 ± 9.11 | 262.31 ± 20.84 | 202.85 ± 10.77 | 288.27 ± 27.64 | 191.88 ± 9.37 | 218.66 ± 19.37 |
| 10 | 491.56 ± 13.31 | 404.69 ± 25.89 | 547.55 ± 21.66 | 573.96 ± 64.47 | 498.72 ± 12.16 | 472.64 ± 36.25 |

**Table 3: Percentage enhancement (%) in heat transfer coefficient associated with bubble-driven particle self-propulsion for the data in**

**Table** 2. Following the "±" are the estimated standard error for each condition (see §S3 of the SI for the formula used).

| Power Input (W) | Rayleigh Number | Percentage Enhancement (%) | | |
|---|---|---|---|---|
| | | Trial 1 | Trial 2 | Trial 3 |
| 0.1 | $1.0 \times 10^4$ | 116.66 ± 50.11 | 62.59 ± 41.58 | 49.17 ± 42.93 |
| 0.4 | $2.0 \times 10^4$ | 52.37 ± 24.85 | 35.52 ± 16.17 | 77.49 ± 24.1 |
| 1 | $3.3 \times 10^4$ | 49.96 ± 14.25 | 42.11 ± 15.58 | 13.96 ± 11.53 |
| 10 | $1.1 \times 10^5$ | -17.67 ± 5.72 | 4.82 ± 12.48 | -5.23 ± 7.63 |

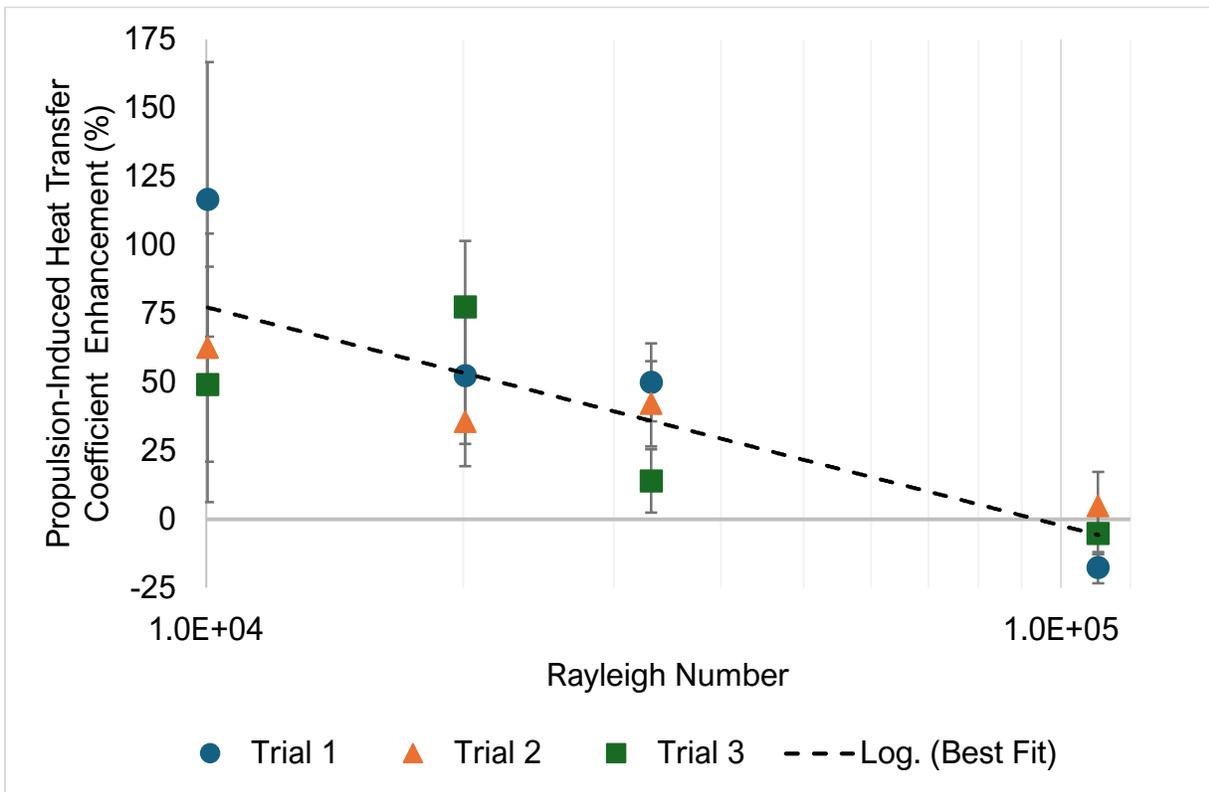

Figure 9: Percentage enhancement in heat transfer as a function of Rayleigh number for the data in Table 3. The equation plotted by the dashed fit line represents a least-squares fit to the mean enhancement at each Rayleigh number, and is given in the main text. Error bars indicate the estimated standard error in each direction (see §S3 of the SI for details on how the standard error was computed).

Table 3 and **Figure 9** show the average percent enhancement in the convective HTC associated with self-propulsion (compared to the non-propelled case) associated with each experimental trial. The dashed line in **Figure 9** represents a logarithmic line of best fit to the mean enhancement percentage at each Rayleigh number, and is described mathematically by the equation $y = -34.6 \ln(Ra_{L_c}) + 395.88$, where $y$ is the percentage enhancement in HTC associated with propulsion compared to no propulsion (that is, $y$ is the ordinate of **Figure 9**). As the heating power (and thus the Rayleigh number) is increased, buoyancy forces on the fluid elements adjacent to the heater surface become stronger relative to viscous forces, and thus flows associated with natural convection heat transfer can become significant. To quantify the relative importance of natural convection, we computed the Rayleigh number, $Ra$, according to equation (2), which is the abscissa of **Figure 9**. In the experiments reported here, the range of $Ra$ is on the order of $10^4$ to $10^5$, which is known to be the range in which natural convection begins to exert a significant influence on the HTC value [51]. Thus, we hypothesize that as the heating power is increased, natural convective flows begin to compete with, and eventually overcome, those generated by the motions of the SPPs and the $O_2$ bubbles.

## V. Discussion

In the first set of experiments, we used infrared (IR) thermography to visualize the influence of self-propelled manganese dioxide ($MnO_2$) microparticles on the temperature distribution of a fluid subjected to a constant heat flux with a vertical heater. While the exact contribution of the SPPs is difficult to disentangle from those of the $O_2$ bubbles they continuously generate, the IR images establish that SPPs generate perturbations to the temperature field that outpace thermal diffusion in the fluid. To quantify thermal mixing enhancement, we then sought to measure the increase to the heat transfer coefficient due to the presence of bubble-generating SPPs.

In the second set of experiments, we demonstrated enhancements of up to 100% in the convective heat transfer coefficient in the case of self-propulsion compared to the case without propulsion. As **Table 3** shows, the enhancement percentage decreases as the heating power is increased. To explain this dependence, we estimated the Rayleigh number for each case, finding values between $10^4$ and $10^5$, which is the range in which natural convective flows begin to overcome viscous dissipation. As the Rayleigh number is increased, natural convective flows become more significant, and we hypothesize that they outcompete and overwhelm the convective enhancements generated by SPPs. Eventually, at high Rayleigh number, propulsion begins to be associated with a slight *reduction* in the HTC compared to the no-propulsion case. While the exact explanation for this is unclear, it is possible that at high Rayleigh number, the SPPs' active motion disrupts the natural convection currents, hindering natural convection and reducing the HTC. Previous designs of artificial SPPs have been known to undergo rheotaxis [57,58], so it is plausible that SPPs undergoing rheotaxis would inhibit natural convection by swimming against the bulk natural convective currents. However, verification of this explanation would require quantitative characterization of the magnitude and direction of the SPPs' flow disturbances (and those of the bubbles), and direct comparison to the velocity fields induced by natural convection. These measurements would yield insight into the interplay between natural convection and SPP-induced convection, which is key to a more rigorous understanding of the results presented herein.

The HTC enhancement results presented in **Figure 7, Figure 8,** and **Table 2** incorporate the effects both the SPPs and the SPP-generated bubbles on thermal mixing in the fluid. It is possible that the rising bubbles themselves play a significant role in the heat transfer enhancement, especially since in this geometry (in which heat is supplied to the fluid from below), the direction of the bubbles' motion is in the same direction as that of heat transfer. Rigorous determination of the relative importance of bubble-induced and SPP-induced thermal mixing is challenging since it would entail gathering the same data for HTC vs. time (as in Figure 7) for an SPP design whose self-propelled speed, direction, shape, size, and volume fraction are identical to bubble-driven $MnO_2$ SPPs, but whose propulsion mechanism does not involve bubbles. Although several SPP designs have been demonstrated, many of which move without generating bubbles, no current design meets these criteria. However, one could qualitatively assess the importance of bubbles by using another design that does not employ bubbles but resembles the $MnO_2$ particles approximately. We have observed similar results using self-propelled grains of camphor roughly 1 mm in size (shown in **Figure S29** in the SI), which suggest that they also can induce similar perturbations in the temperature field. However, these designs rely on the asymmetric dissolution of camphor, and the Marangoni forces that result from this asymmetry, and so are not practical to use in the current experimental setup. Future work should aim to replicate these experiments but with an SPP design that moves efficiently, within the 3-dimensional bulk fluid, and without generating bubbles.

Another possible contribution to consider is the heat released during hydrogen peroxide decomposition, which is exothermic ($\Delta G° = -224$ kJ/mol). To ensure that the enhancement in thermal mixing was not an artifact of this exothermicity, a control experiment was conducted using the same Petri dish setup pictured in Figure 4, with the same conditions as the primary experiments, ($H_2O_2$ 0.3 wt. %, $MnO_2$ 1 mg/mL), only lacking surfactant, which prevented bubble detachment (and thus propulsion). Under these conditions, with the Peltier heater turned off, a negligible temperature rise was observed in the $MnO_2/H_2O_2$ suspension over the course of 10 minutes, suggesting that the exothermicity of $H_2O_2$ decomposition did not significantly affect the results.

Finally, in this work we do not directly address the contributions of heat conduction through the interior of the $MnO_2$ particles themselves. This is because our goal in this work was to glean insight into the effect of particle self-propulsion alone. Hence, the control for each trial is a suspension of the exact same particles but without fuel (and thus no propulsion), meaning that the conductive heat transfer rate through the particles was kept the same in the self-propelled and control cases. Since (as **Table 1** shows) the thermal conductivity of aqueous $H_2O_2$ solutions is consistently lower than that of pure water, it can be inferred that the enhancement in the heat transfer coefficient did not result from any improvement in the thermal conductivity of the solution.

## *VI.    Conclusion*

This work has provided experimental data quantifying the enhancement of convective heat transfer by self-propelled particles (i.e., those that do not require external forcing). Our results suggest that the presence of bubble-propelled SPPs can outpace thermal diffusion and enhance heat transfer relative to the same suspension without fuel.

A logical follow-up to this work would be to conduct thorough study quantifying the effect of particle size, morphology, volume fraction, velocity, rotation, and method of propulsion on the advective potential of SPPs. Second, an investigation into the behavior of SPPs under various convective regimes, such as natural convection at higher Rayleigh numbers or forced convection at various Reynolds numbers. Such studies could utilize particle image velocimetry (PIV) of the SPP-induced flow fields coupled to the temperature field to better understand the particle behavior. Third, an examination of the rheological behavior of the SPP suspension under various conditions could lend insight into how the swimmers are changing the viscosity of the fluid. Lastly, future work could attempt to optimize the trajectories of the microparticles by changing the parameters that affect propulsion in order to induce active turbulence in the fluid. This could be done with various swimmer designs, including both those with and without bubble generation.

Another promising avenue for future work would be to quantify the changes in viscosity resulting from the SPPs' motion. It has already been shown that the disturbance flows from biological microswimmers can raise or lower the viscosity of the suspension, relative to the liquid alone, depending on the swimming strategy. Specifically, "pushers" (propelled from behind) can induce a net reduction in the suspension's viscosity compared to the liquid alone [38], so much so that "superfluid"-like behavior has been reported [36,59] [60], in which the suspension flows in the absence of an applied shear. In contrast to pushers, pullers do not generate significant chaotic flows, and lead to an overall increase in the suspension viscosity (as passive particles do) [32,38]. The rich rheological behavior of active suspensions was reviewed by Saintillan [38]. Considering the significant disadvantages associated with the increases in viscosity of conventional nanofluids [24,25,61], an SPP suspension exhibiting both enhanced thermal properties and reduced viscosity relative to the liquid alone would be appealing.

A few potential applications of this work extend from microelectronic devices to large-scale industrial data centers. SPPs could be employed to enhance heat transfer in setups that are difficult to cool with pumps. SPPs could also be used to remove large temperature gradients near the heat source within the thermal boundary layer.

## *Acknowledgments*

This work was supported by the US National Science Foundation grants CBET-2039262 and CBET-2039263 (to J.L.M. and P.K., respectively) and the George Mason University Office of Student Scholarship, Creative Activities, and Research (OSCAR) via an Undergraduate Research Scholars Program (URSP) fellowship to J.B.V. The authors thank Dr. K. Lee (director of OSCAR) for support. The authors also thank Prof. A. Kr. Singh (Woxsen University), N. Lechner (George Mason U.), S. Kargar (George Mason U.), and D. Devkar (U. Virginia) for their assistance and advice. The authors thank Prof. M. Bucci (MIT), Prof. C. Markides (Imperial College London), and Dr. Y. El Hasadi (TU Delft) for fruitful discussions.## *References*

1. Jones N. How to stop data centres from gobbling up the world's electricity. Nature. 2018 Sep 12;561(7722):163–6.

2. Belkhir L, Elmeligi A. Assessing ICT global emissions footprint: Trends to 2040 & recommendations. J Clean Prod. 2018 Mar 10;177:448–63.

3. Electric Power Research Institute. Powering Intelligence: Analyzing Artificial Intelligence and Data Center Energy Consumption [Internet]. 2024 [cited 2024 Jul 16]. Available from: https://www.epri.com/research/products/3002028905

4. Coles H, Herrlin M. Immersion Cooling of Electronics in DoD Installations. CALIFORNIA UNIV BERKELEY BERKELEY United States; 2016.

5. Choi SUS, Eastman JA. Enhancing thermal conductivity of fluids with nanoparticles [Internet]. Argonne National Lab., IL (United States); 1995 Oct [cited 2019 Mar 25]. Report No.: ANL/MSD/CP-84938; CONF-951135-29. Available from: https://www.osti.gov/biblio/196525-enhancing-thermal-conductivity-fluids-nanoparticles

6. Eastman JA, Choi SUS, Li S, Yu W, Thompson LJ. Anomalously increased effective thermal conductivities of ethylene glycol-based nanofluids containing copper nanoparticles. Appl Phys Lett. 2001 Feb 2;78(6):718–20.

7. Choi SUS, Zhang ZG, Yu W, Lockwood FE, Grulke EA. Anomalous thermal conductivity enhancement in nanotube suspensions. Appl Phys Lett. 2001 Sep 24;79(14):2252–4.

8. Prasher R, Bhattacharya P, Phelan PE. Thermal Conductivity of Nanoscale Colloidal Solutions (Nanofluids). Phys Rev Lett. 2005 Jan 18;94(2):025901.

9. Babaei H, Keblinski P, Khodadadi JM. A proof for insignificant effect of Brownian motion-induced micro-convection on thermal conductivity of nanofluids by utilizing molecular dynamics simulations. J Appl Phys. 2013 Feb 22;113(8):084302.

10. Evans W, Fish J, Keblinski P. Role of Brownian motion hydrodynamics on nanofluid thermal conductivity. Appl Phys Lett. 2006 Feb 27;88(9):093116.

11. Eapen J, Williams WC, Buongiorno J, Hu L wen, Yip S, Rusconi R, Piazza R. Mean-Field Versus Microconvection Effects in Nanofluid Thermal Conduction. Phys Rev Lett. 2007 Aug 28;99(9):095901.

12. Putnam SA, Cahill DG, Braun PV, Ge Z, Shimmin RG. Thermal conductivity of nanoparticle suspensions. J Appl Phys. 2006 Apr 15;99(8):084308.
31

13. Buongiorno J, Venerus DC, Prabhat N, McKrell T, Townsend J, Christianson R, Tolmachev YV, Keblinski P. A benchmark study on the thermal conductivity of nanofluids. J Appl Phys. 2009 Nov 1;106(9):094312.

14. Maxwell JC. A treatise on electricity and magnetism. 2nd ed. Oxford: Clarendon Press; 1881.

15. Nan CW, Birringer R, Clarke DR, Gleiter H. Effective thermal conductivity of particulate composites with interfacial thermal resistance. J Appl Phys. 1997 May 15;81(10):6692–9.

16. Prabhat N, Buongiorno J, Hu LW. Convective heat transfer enhancement in nanofluids: real anomaly or analysis artifact? J Nanofluids. 2012;1(1):55–62.

17. Keblinski P, Prasher R, Eapen J. Thermal conductance of nanofluids: is the controversy over? J Nanoparticle Res. 2008 Oct 1;10(7):1089–97.

18. Wang XQ, Mujumdar AS. Heat transfer characteristics of nanofluids: a review. Int J Therm Sci. 2007 Jan 1;46(1):1–19.

19. Angayarkanni SA, Philip J. Review on thermal properties of nanofluids: Recent developments. Adv Colloid Interface Sci. 2015 Nov 1;225:146–76.

20. Evans W, Prasher R, Fish J, Meakin P, Phelan P, Keblinski P. Effect of aggregation and interfacial thermal resistance on thermal conductivity of nanocomposites and colloidal nanofluids. Int J Heat Mass Transf. 2008 Mar 1;51(5):1431–8.

21. Foygel M, Morris RD, Anez D, French S, Sobolev VL. Theoretical and computational studies of carbon nanotube composites and suspensions: Electrical and thermal conductivity. Phys Rev B. 2005 Mar 4;71(10):104201.

22. Keblinski P, Phillpot SR, Choi SUS, Eastman JA. Mechanisms of heat flow in suspensions of nano-sized particles (nanofluids). Int J Heat Mass Transf. 2002 Feb 1;45(4):855–63.

23. Philip J, Shima PD, Raj B. Evidence for enhanced thermal conduction through percolating structures in nanofluids. Nanotechnology. 2008 Jun;19(30):305706.

24. Duan F, Kwek D, Crivoi A. Viscosity affected by nanoparticle aggregation in Al2O3-water nanofluids. Nanoscale Res Lett. 2011 Mar 22;6(1):248.

25. Bashirnezhad K, Bazri S, Safaei MR, Goodarzi M, Dahari M, Mahian O, Dalkılıça AS, Wongwises S. Viscosity of nanofluids: A review of recent experimental studies. Int Commun Heat Mass Transf. 2016 Apr 1;73:114–23.

26. Buongiorno J. Convective Transport in Nanofluids. J Heat Transf. 2006 Mar 1;128(3):240–50.

27. Paxton WF, Kistler KC, Olmeda CC, Sen A, St Angelo SK, Cao YY, Mallouk TE, Lammert PE, Crespi VH. Catalytic nanomotors: Autonomous movement of striped nanorods. J Am Chem Soc. 2004 Oct 20;126(41):13424–31.

28. Chen X, Zhou C, Wang W. Colloidal Motors 101: A Beginner's Guide to Colloidal Motor Research. Chem – Asian J. 14(14):2388–405.

29. Lauga E, Powers TR. The hydrodynamics of swimming microorganisms. Rep Prog Phys. 2009;72(9):096601.

30. Wu XL, Libchaber A. Particle Diffusion in a Quasi-Two-Dimensional Bacterial Bath. Phys Rev Lett. 2000 Mar 27;84(13):3017–20.

31. Kim MJ, Breuer KS. Enhanced diffusion due to motile bacteria. Phys Fluids. 2004 Aug 6;16(9):L78–81.

32. Saintillan D, Shelley MJ. Emergence of coherent structures and large-scale flows in motile suspensions. J R Soc Interface. 2012;9:571–85.

33. Saintillan D, Shelley MJ. Instabilities and Pattern Formation in Active Particle Suspensions: Kinetic Theory and Continuum Simulations. Phys Rev Lett. 2008 Apr 29;100(17):178103.

34. Dunkel J, Heidenreich S, Drescher K, Wensink HH, Bär M, Goldstein RE. Fluid Dynamics of Bacterial Turbulence. Phys Rev Lett. 2013 May 28;110(22):228102.

35. Gachelin J, Miño G, Berthet H, Lindner A, Rousselet A, Clément É. Non-Newtonian Viscosity of Escherichia coli Suspensions. Phys Rev Lett. 2013 Jun 26;110(26):268103.

36. López HM, Gachelin J, Douarche C, Auradou H, Clément E. Turning Bacteria Suspensions into Superfluids. Phys Rev Lett. 2015 Jul 7;115(2):028301.

37. Rafaï S, Jibuti L, Peyla P. Effective Viscosity of Microswimmer Suspensions. Phys Rev Lett. 2010 Mar 3;104(9):098102.

38. Saintillan D. Rheology of Active Fluids. Annu Rev Fluid Mech. 2018;50(1):563–92.

39. Solis KJ, Martin JE. Multiaxial fields drive the thermal conductivity switching of a magneto-responsive platelet suspension. Soft Matter. 2013 Sep 12;9(38):9182–8.

40. Sommer T, Danza F, Berg J, Sengupta A, Constantinescu G, Tokyay T, Bürgmann H, Dressler Y, Steiner OS, Schubert CJ, Tonolla M, Wüest A. Bacteria-induced mixing in natural waters. Geophys Res Lett. 2017;44(18):9424–32.

41. Wang Z, Mathai V, Sun C. Self-sustained biphasic catalytic particle turbulence. Nat Commun. 2019 Jul 26;10(1):3333.

42. Wang Z, Mathai V, Sun C. Experimental study of the heat transfer properties of self-sustained biphasic thermally driven turbulence. Int J Heat Mass Transf. 2020 May 1;152:119515.

43. El Hasadi YMF, Crapper M. Self-propelled nanofluids a path to a highly effective coolant. Appl Therm Eng. 2017 Dec 25;127:857–69.

44. El Hasadi YMF, Crapper M. Self-propelled nanofluids a coolant inspired from nature with enhanced thermal transport properties. J Mol Liq. 2020 Jun 12;113548.

45. Peng W, Chandra A, Keblinski P, Moran JL. Thermal transport dynamics in active heat transfer fluids (AHTF). J Appl Phys. 2021 May 6;129(17):174702.

46. Wang H, Zhao G, Pumera M. Beyond Platinum: Bubble-Propelled Micromotors Based on Ag and MnO2 Catalysts. J Am Chem Soc. 2014 Feb 19;136(7):2719–22.

47. Soler L, Magdanz V, Fomin VM, Sanchez S, Schmidt OG. Self-Propelled Micromotors for Cleaning Polluted Water. ACS Nano. 2013 Nov 26;7(11):9611–20.

48. Moran JL, Posner JD. Phoretic Self-Propulsion. Annu Rev Fluid Mech. 2017;49(1):511–40.

49. Bailey MR, Fedosov DA, Paratore F, Grillo F, Gompper G, Isa L. Low efficiency of Janus microswimmers as hydrodynamic mixers. Phys Rev E. 2024 Oct 1;110(4):044601.

50. Wang H, Zhao G, Pumera M. Crucial Role of Surfactants in Bubble-Propelled Microengines. J Phys Chem C. 2014 Mar 13;118(10):5268–74.

51. Bergman TL, Incropera FP, DeWitt DP, Lavine AS. Fundamentals of heat and mass transfer. John Wiley & Sons; 2011.

52. Robinson PJ, Davies JA. Laboratory Determinations of Water Surface Emissivity. 1972 Dec 1 [cited 2025 Aug 11]; Available from: https://journals.ametsoc.org/view/journals/apme/11/8/1520-0450_1972_011_1391_ldowse_2_0_co_2.xml

53. Shaw JA, Cimini D, Westwater ER, Han Y, Zorn HM, Churnside JH. Scanning infrared radiometer for measuring the air–sea temperature difference. Appl Opt. 2001 Sep 20;40(27):4807–15.

54. ToolBox TE. Water - Thermal Diffusivity vs. Temperature and Pressure [Internet]. 2018 [cited 2025 Jun 25]. Available from: https://www.engineeringtoolbox.com/water-steam-thermal-diffusivity-d_2058.html

55. ToolBox TE. Water - Dynamic and Kinematic Viscosity at Various Temperatures and Pressures [Internet]. 2018 [cited 2025 Jun 25]. Available from: https://www.engineeringtoolbox.com/water-dynamic-kinematic-viscosity-d_596.html

56. Takatori SC, Brady JF. Forces, stresses and the (thermo?) dynamics of active matter. Curr Opin Colloid Interface Sci. 2016 Feb 1;21:24–33.

57. Brosseau Q, Usabiaga FB, Lushi E, Wu Y, Ristroph L, Zhang J, Ward M, Shelley MJ. Relating Rheotaxis and Hydrodynamic Actuation using Asymmetric Gold-Platinum Phoretic Rods. Phys Rev Lett. 2019 Oct 25;123(17):178004.

58. Palacci J, Sacanna S, Abramian A, Barral J, Hanson K, Grosberg AY, Pine DJ, Chaikin PM. Artificial rheotaxis. Sci Adv. 2015 May 1;1(4):e1400214–e1400214.

59. Takatori SC, Brady JF. Superfluid Behavior of Active Suspensions from Diffusive Stretching. Phys Rev Lett. 2017 Jan 6;118(1):018003.

60. Słomka J, Dunkel J. Spontaneous mirror-symmetry breaking induces inverse energy cascade in 3D active fluids. Proc Natl Acad Sci. 2017 Feb 28;114(9):2119–24.

61. Kwak K, Kim C. Viscosity and thermal conductivity of copper oxide nanofluid dispersed in ethylene glycol. Korea-Aust Rheol J. 2005;17(2):35–40.